\documentclass[12pt,a4paper ]{article}

\usepackage{graphicx}

\input epsf

\newcommand{\frat}[2]{\frac{\textstyle #1}{\textstyle #2}}

\newcommand{\dmn}[2]{\mbox{$#1\!\cdot\! 10^{#2}\,$}}

\newcommand{\nomer}[1]{\mbox{$\cal N$\hspace{-.5ex}\raisebox{.3ex}
{\underline{\tiny 0}$\!$} #1}}

\begin{document}
\begin{center}
{\Large \bf Towards self-consistent definition of instanton liquid parameters}\\
 \vspace{0.5cm} S.V. Molodtsov$^{1,2}$,  G.M. Zinovjev$^{3}$
\\ \vspace{0.5cm}
{\small $^1$Joint Institute for Nuclear Research, RU-141980,
Dubna, Moscow region, Russia}
\\ \vspace{0.5cm} {\small $^2$Institute of Theoretical and Experimental Physics,
RU-117259, Moscow, Russia}
\\ \vspace{0.5cm} {\small $^3$Bogolyubov Institute for Theoretical Physics,
National Academy of Sciences of Ukraine, UA-03680, Kiev-143, Ukraine}
\end{center}
\vspace{0.5cm}

\begin{center}
\begin{tabular}{p{16cm}}
{\small{The possibility of self-consistent determination of instanton liquid parameters
is discussed together with the definition of optimal pseudo-particle configurations and
comparing the various pseudo-particle ensembles. The weakening of repulsive interactions
between pseudo-particles is
argued and estimated.}}
\end{tabular}
\end{center}
\vspace{0.5cm}

The problem of finding the most effective pseudo-particle profile for instanton liquid (IL)
model of the QCD vacuum \cite{1} has already been formulated in the first papers treating
the pseudo-particle superposition as the quasi-classical configuration saturating the generating
functional \cite{2} of the following form
\begin{equation}
\label{1}
Z=\int D[ {\cal  A}]~e^{-S( {\cal  A})}~,
\end{equation}
where $S( {\cal A})$ is the Yang-Mills action. Although the solution proposed in Ref. \cite{2}
was quite acceptable phenomenologically the consequent more accurate analysis discovered several
imperfect conclusions putting into doubt the assertion about the instanton ensemble getting stabilization
and some additional mechanism should be introduced to fix such an ensemble \cite{3}.
In this note we revisit the task formulated in Ref. \cite{2} within the self-consistent approach proposed
in our previous paper \cite{4}. We are not speculating on the detailed mechanism of stabilizing and are
based on one crucial assumption which is the existence of non-zero gluon condensate in the QCD vacuum.
This idea is not very original but turns out far reaching in the context of our approach.
The particular form and properties of this condensate will be discussed in the following paper.

Thus, as the configuration saturating the generating functional (\ref{1}) we take the following superposition
\begin{equation}
\label{2}
{\cal  A}^{a}_\mu(x)=B^a_{\mu}(x)+\sum_{i=1}^N
A^{a}_\mu(x;\gamma_i)~,
\end{equation}
here $A^a_{\mu}$ stands for the (anti-)instanton field in the singular gauge
\begin{equation}
\label{3}
A^a_{\mu}(x;\gamma)=\frat2g~\omega^{ab}\bar\eta_{b\mu\nu}~\frat{y_\nu}{y^2}~f(y),~~~y=x-z~,
\end{equation}
$\gamma_i=(\rho_i,z_i,\omega_i)$ denotes all the parameters describing the
$i$-th (anti-)instanton, in particular, its size $\rho$, colour orientation $\omega$,
center position $z$ and as usual $g$ is the coupling constant of gauge field.
The function $f(y)$ introduces the pseudo-particle profile and will be fixed by resolving
the suitable variational problem. For example, for the conventional singular instanton it looks like
\begin{equation}
\label{4}
f(y)=\frat{1}{1+\frat{y^2}{\rho^2}}~.
\end{equation}
In analogy with this form we consider the function $f$ depending on $y^2$ or,
more precisely, on the variable ${\mbox{x}}=\frat{y^2}{\bar\rho^2}$ at some characteristic
mean pseudo-particle size $\bar\rho$. Dealing with the anti-instanton one should make the
substitution of the 't Hooft symbol $\bar\eta \to \eta$. It is seen from (\ref{2}) we 'singled out'
one pseudo-particle of ensemble and introduced the special symbol $B$ for its field which actually
has the same form as Eq. (\ref{3}).

The strength tensor of this 'external' field and the field of every separate pseudo-particle $A$ can be written as
\begin{equation}
\label{5}
G_{\mu\nu}^a=G_{\mu\nu}^a(B)+G_{\mu\nu}^a(A)+G_{\mu\nu}^a(A,B)~,
\end{equation}
where two first terms are given by the standard definition of field strength
\begin{equation}
\label{6}
 G_{\mu\nu}^a(A)=\partial_\mu  A^a_{\nu}-\partial_\nu
 A^a_{\mu}+g~f^{abc} A^b_{\mu} A^c_{\nu}~,
\end{equation}
with the entirely antisymmetric tensor $f^{abc}$. In particular, for the singular instanton of
Eq. (\ref{3}) it takes the form
\begin{equation}
\label{7}
G^a_{\mu\nu}=-\frat4g~\omega^{ak}\left[\bar\eta_{k\alpha\beta}
~\frat{f(1-f)}{y^2}+(\bar\eta_{k\mu\beta}~y_\nu-\bar\eta_{k\nu\alpha}~y_\mu)~\frat{y_\alpha}{y^2}~
\left(f'-\frat{f(1-f)}{y^2}\right)\right]~,
\end{equation}
where $f'$ means the derivative over $y^2$. The third term of Eq. (\ref{5}) presents the 'mixed'
component of field strength and is
\begin{equation}
\label{8}
G_{\mu\nu}^a(A,B)=g~f^{abc}(B^b_{\mu}A^{c}_\nu-B^b_{\nu}A^{c}_\mu)
=g~f^{abc}\omega^{cd}~\frat2g~ (B^{b}_\mu
~\bar\eta_{d\nu\alpha}-B^{b}_\nu
~\bar\eta_{d\mu\alpha})~\frat{y_{\alpha}}{y^2}~f.
\end{equation}

It was shown in Ref. \cite{4} that in quasi-classical regime which is of particular interest for
applications, the generating functional (\ref{1}) could be essentially simplified if reformulated
in terms of the field $B_{\cal A}$ averaged over ensemble ${\cal  A}$. Performing the cluster
decomposition \cite{5} of stochastic exponent in Eq. (\ref{1})
\begin{equation}
\label{9}
\langle \exp (-S)\rangle_{\omega z}=\exp\left(~\sum_k
\frat{(-1)^k}{k!}~\langle\langle S^k\rangle\rangle_{\omega z}\right)~,
\end{equation}
where $\langle S_1\rangle=\langle\langle S_1\rangle\rangle$,
$\langle S_1 S_2\rangle=\langle S_1\rangle\langle S_2\rangle
+\langle\langle S_1 S_2\rangle\rangle, \dots$ (the first cumulant is simply defined by averaging the action)
the higher terms of effective action for the 'external' field in IL could be presented as
\begin{equation}
\label{10}
\langle \langle S[B_{\cal  A}] \rangle\rangle_{\cal  A}=\int d^4x~
\left( \frat{G(B_{\cal  A})~G(B_{\cal  A})}{4}+\frat{m^2}{2}~B_{{\cal  A}}^2\right)~,
\end{equation}
and the mass $m$ is defined by the IL parameters developing for the standard singular
pseudo-particles (\ref{4}) the following form (see, also below)
\begin{equation}
\label{11}
m^2=9\pi^2~n~\bar\rho^2~\frat{N_c}{N_c^{2}-1}~,
\end{equation}
with $n=N/V$ where $N$ is the total number of pseudoparticles in the volume $V$ and $N_c$ is the number of colours.
The small magnitude of characteristic IL parameter (packing fraction) $n\bar\rho^4$ allows us at decomposing
to keep the contributions of one pseudo-particle term ($\sim n$) only.

The effective action in Eq. (\ref{10}) implies a functional integration in which the vacuum stochastic
fields are not destroyed by the external field. Then there is no reason to develop the detailed
description of the field $B$ driven by the symmetries of initial gauge invariant Lagrangian for the
Yang-Mills fields. In practice it could be understood as an argument to do use the averaged action
dealing with the field $B$. It means the colourless binary (and similar even) configurations only
of field $B$ survive in the effective action.
In other words the decomposition $B\simeq B_{\cal A}+\cdots$ is used (in what follows we are not maintaining
the index for the field $B$). Obviously, if there is any need of more detailed description including, for example,
information on the fluctuations of field $B$ one should operate with the correlation functions
of higher order and the corresponding chain of the Bogolyubov equations.

The selfconsistent description of pseudo-particle ensemble may not be developed based on
Eq. (\ref{10}) only because in such a form the pseudo-particles of zero size $\rho=0$
are most advantageous. In Ref. \cite{4} the version of variational principle was proposed
which makes it possible to determine the selfconsistent solution in long wave-length approximation
for the pseudo-particle ensemble (anti-instantons in the singular gauge with standard
profile (\ref{4})) and external field. Here it adapts to the saturating configuration (\ref{2})
also and its more optimal (than standard) profile is defined, as suggested in Ref. \cite{2},
taking into account the IL parameter change while the pseudo-particle field is present.

The contribution of saturating configuration into the generating functional is evaluated as
(see \cite{2} for the denotions)
\begin{equation}
\label{12}
Z\simeq Y=\int D[B]~ \frat{1}{N!} \int \prod_{i=1}^N~ d\gamma_i~~e^{-S(B,\gamma)}~.
\end{equation}
The following terms should be taken into consideration
\begin{equation}
\label{13}
S(B,\gamma)=-\sum_{i=1}^N \ln d(\rho_i)+ \beta~U_{int}+\sum_{i=1}^N U_{ext}^i(B)+S(B)~,
\end{equation}
(the details of deducing this expression can be found in \cite{4}).
Here we remind only that to obtain it one should average over the pseudo-particle
parameters and to hold the highest contributions only at summing up the pseudo-particles.
If the saturating configurations are the instantons in singular gauge with the standard profile (\ref{4})
the first term describing the one instanton contributions takes the form of distribution function over
(anti-)instanton sizes
\begin{equation}
\label{14} d(\rho)= C_{N_c} \Lambda^b~\rho^{b-5}
\widetilde\beta^{2 N_c},
\end{equation}
where
\begin{equation}
\label{15}
b=\frac{11}{3} N_c-\frac{2}{3}N_f~,
\end{equation}
$\widetilde\beta=-b \ln(\Lambda \bar\rho)$,
$$C_{N_c}\approx\frac{4.66~\exp(-1.68 N_c)}{\pi^2(N_c-1)!(N_c-2)!}~.$$
If one considers the profile of Eq. (\ref{3}) the change of one pseudo-particle action which has the form
\begin{equation}
\label{16}
S_i=3~\int_0^{\infty}\frat{d y^2}{y^2}~\beta~\left[(y^2 f')^2+f^2(1-f)^2\right]~,
\end{equation}
should be absorbed while calculating. Here $\beta=8\pi^2/g^2$ is the characteristic action of single
pseudo-particle (\ref{4}) which is defined at the scale of average pseudo-particle size
$\beta=\beta(\bar\rho)$ where $\beta(\rho)=-\ln C_{N_c}-b \ln(\Lambda \rho)$. The coefficient
$b$ enters the corresponding equations (in particular the distribution function (\ref{14}))
always with the additional factor $s=\frat{S_i}{\beta}$. It means that in all the formula
containing the one instanton contribution the following substitution
\begin{equation}
\label{17}
b\to b~s~.
\end{equation}
should be done. The penultimate term of Eq. (\ref{13}) accumulates the partial pseudo-particle
contributions coming from the 'mixed' component of the strength tensor (\ref{8}) and describing
the interaction of pseudo-particle ensemble with the detached one, i.e.
$$U_{ext}^i(B)=\int d^4x ~\left\langle\frat{G_{\mu\nu}^a(A_i,B)~
G_{\mu\nu}^a(A_i,B)}{4}\right\rangle_{\gamma_i}~.$$
The other terms at the characteristic IL parameters are small as it was shown in Ref. \cite{4}.
The average value of 'mixed' component is given by the following formula
\begin{equation}
\label{18}
\langle G_{\mu\nu}^a(A,B)~G_{\mu\nu}^a(A,B)
\rangle_{\omega z}= \frat{18}{V}\frat{N_c}{N_c^{2}-1}~I
~~B^{b}_\mu ~B^{b}_\mu~,~~B^2=\frat{12}{g^2}~\frat{f^2}{y^2}~,
\end{equation}
here $I$ is defined by the integrated profile function of pseudo-particle
$$I_{\alpha,\beta}=\delta_{\alpha,\beta}~I=\int dy~ \frat{y_\alpha y_\beta}{y^4}~f^2~,~~
I=\frat{\pi^2\rho^2}{4}~\int_{0}^\infty d{\mbox{x}}~f^2~,~~{\mbox{x}}=\frat{y^2}{\rho^2}~.
$$
In particular, for the standard form of pseudo-particle we have
$$\int_{0}^\infty d{\mbox{x}}~f^2=1~.
$$
The corresponding constant (see \cite{4}) $\zeta_0=\frac{9~\pi^2}{2}~\frac{N_c}{N_c^2-1}$ should
be changed for the modified one
$$\zeta=\lambda \zeta_0~,~~{\mbox{ё}}~~\lambda=\int_{0}^\infty d{\mbox{x}}~f^2~,$$
in all terms describing the interaction of IL with detached pseudo-particle if the profile
function $f$ is arbitrary. Eq. (\ref{18}) demonstrates that we are formally dealing with non-zero
value of gluon condensate which is given by the correlation function
\begin{equation}
\label{19}
\langle A^a_{\mu}(x;\gamma) A^a_{\mu}(y;\gamma) \rangle_{\omega z}=
\frat{4}{g^2}~\frat{N_c}{N_c^{2}-1}\frat{\rho^2}{V}~
F\left(\frat{|x-y|}{\rho}\right)~.
\end{equation}
For the pseudo-particle of standard form the function $F(\Delta)$ equals to
\begin{eqnarray}
\label{20}
F(\Delta)&=&\frat{\pi^2}{4}~\frat{\Delta^2+2}{|\Delta|}\sqrt{\Delta^2+4}~
\ln\left|\frat{\sqrt{\Delta^2+4}(\Delta^2+1)+\Delta^3+3\Delta}
{\sqrt{\Delta^2+4}-\Delta}\right|-\nonumber\\ [-.2cm]
\\[-.25cm]
&-&\pi^2~\frat{(\Delta^2+1)^2}{\Delta^2}~\ln(1+\Delta^2)+\pi^2~\Delta^2~\ln
|\Delta|~,\nonumber
\end{eqnarray}
with the asymptotic behaviours
$$\lim_{\Delta\to 0} F(\Delta)\to \pi^2-\frat{\pi^2}{3}~
\Delta^2+\pi^2 ~\Delta^2~\ln |\Delta|~,~~~~~\lim_{\Delta\to
\infty} F(\Delta)\to \frat{\pi^2}{\Delta^2}~.$$
The presence of this condensate (\ref{19}) which leads, in particular, to the mass definition as in
(\ref{11}) just signifies the assumption mentioned at the beginning this note.

The second term of (\ref{13}) describes the repulsive interaction between the pseudo-particles of ensemble
$$\beta~U_{int}=\sum_{i,j}\int d^4x ~\left\langle\frat{G_{\mu\nu}^a(A_i,A_j)~
G_{\mu\nu}^a(A_i,A_j)}{4}\right\rangle_{\gamma_i, \gamma_j}~,
$$
and actually presents the same contribution as $U_{ext}$ but being integrated with the field $B$ of
every individual pseudo-particle as $\beta~U_{int}=\int d^4x ~\frat{m^2}{2}~B^2$. It results in the
change of coupling constant $\xi_{0}^2=\frac{27~\pi^2}{4}\frac{N_c}{N_c^{2}-1}$ describing the
pseudo-particle interaction (see \cite{2}) for new form
$$\xi^2=\lambda^2~\xi_{0}^2~,$$
(similar to the change of constant $\zeta$). And eventually the last term of Eq. (\ref{13}) presents
simply the Yang-Mills action of the $B$ field
$$S(B)=\int d^4x~\frat{G_{\mu\nu}^a(B)~ G_{\mu\nu}^a(B)}{4}~.$$
It is worthwhile to notice that the topological charge of the configuration (\ref{4}) is retained to be equal to
$$N=\frat{1}{\beta}~\int d^4x~ \frat{G^a_{\mu\nu}\widetilde G^a_{\mu\nu}}{4}=-6~
\int_{0}^\infty d{\mbox{x}}~f' f(1-f)=1~,~~
\widetilde G^a_{\mu\nu}=\frat{1}{2}~\varepsilon_{\mu\nu\alpha\beta}~G^a_{\alpha\beta}~,
$$
here $\varepsilon_{\mu\nu\alpha\beta}$ is an entirely antisymmetric tensor, $\varepsilon_{1234}=1$.

The generating functional (\ref{12}) might be estimated with the approximating functional (see \cite{2}) as
\begin{equation}
\label{21} Y\ge Y_1~\exp(-\langle S-S_1\rangle)~,
\end{equation}
where
$$Y_1=\int D[B]~ \frat{1}{N!} \int \prod_{i=1}^N~ d\gamma_i~~e^{-S_1(B,\gamma)-S(B)}~,
~~~S_1(B,\gamma)=-\sum \ln \mu(\rho_i)~,$$
and $\mu(\rho)$ is an effective one particle distribution function defined by solving the variational problem.
In our particular situation the average value of difference of the actions is given as follows
\begin{eqnarray}
\label{22}
&&\langle S-S_1\rangle=\frat{1}{Y_1}~\frat{1}{N!} \int~\prod_{i=1}^N~ d
\gamma_i~ [\beta~U_{int}+U_{ext}(\gamma,B)-\sum\ln d(\rho_i)+\sum \ln \mu(\rho_i)]
~e^{~\sum \ln \mu (\rho_i)}=\nonumber\\
&&=\frat{N}{\mu_0}~\int d \rho~ \mu
(\rho)~\ln \frat{\mu (\rho)}{d (\rho)}+\frat{\beta}{2}~\frat{N^2}{V^2}~\frat{1}{\mu_0^{2}} ~\int d\gamma_1
d\gamma_2~U_{int}(\gamma_1,\gamma_2)~ \mu (\rho_1)
 \mu (\rho_2)+\nonumber\\
 &&+\int d^4x~ \frat{N}{V}~\int d \rho~\frat{\mu (\rho)}{\mu_0}~\rho^2
 \zeta~B^2=\nonumber\\
&&=\int d^4x~ n~ \left(~\int d \rho~ \frat{\mu (\rho)}{\mu_0}~\ln
\frat{\mu (\rho)}{d (\rho)}+\frat{\beta \xi^2}{2}~n \left(
\overline{ \rho^2}\right)^2 +\zeta \overline{\rho^2}~B^2\right)~,
\end{eqnarray}
with $\mu_0=\int d \rho~ \mu (\rho)$. In this note we estimate the functionals in the long wave length
(adiabatic) approximation, i.e. consider the IL elements to be equilibrated by the external fixed field $B$.
Afterwards, with finding the optimal IL parameters out we receive the effective action for the external field
in the selfconsistent form. Eq. (\ref{22}) is taken just in such a form in order to underline the integration
is executed over the IL elements and the parameters describing their states are the functions of external field
(i.e. could finally be the functions of a coordinate ${\mbox{x}}$). The physical meaning of such a functional
is quite transparent and implies that each separate IL element develops its characteristic screening of
the attached field.

\begin{figure*}[!tbh]
\begin{center}
\includegraphics[width=0.5\textwidth]{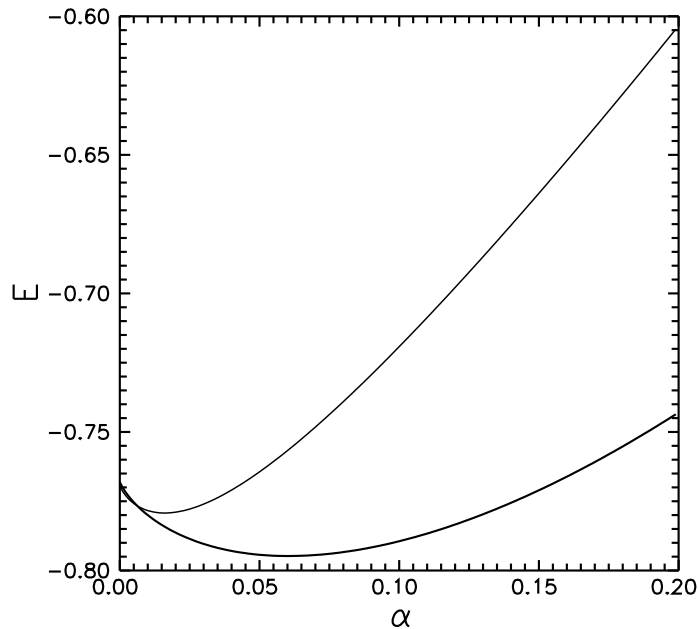}
\end{center}
  \vspace{-7mm}
 \caption{The energy $E(\alpha)$ when the profile function includes a screening effect
(\ref{29}) with the parameter $\lambda$ ($s=1$) only taken into consideration (lower curve)
and with both parameters used (upper curve) (see the text).}
  \label{fig1}
\end{figure*}

Now calculating the variation of action difference $\langle S-S_1\rangle$ over $\mu (\rho)$ we obtain
$$\mu (\rho)= C~ d (\rho) ~e^{-(n\beta\xi^2\overline{\rho^2}+\zeta B^2)\rho^2}~,
$$
where $C$ is an arbitrary constant and its value is fixed by requiring the coincidence of
the distribution function when the external field is switched off ($B=0$) with vacuum distribution function then
\begin{equation}
\label{23}
\mu (\rho)= C_{N_c} \widetilde \beta^{2N_c}\Lambda^{b s} \rho^{b s-5}
 ~e^{-(n\beta\xi^2\overline{\rho^2}+\zeta B^2)\rho^2}~.
\end{equation}
With defining the average size as
$$\overline{\rho^2}=\frat{\int d \rho~ \rho^2~\mu (\rho)}{\mu_0}~,
$$
we come to the practical interrelation between the IL density and average size of pseudo-particles
\begin{equation}
\label{24}
(n~\beta~\xi^2~\overline{\rho^2}+\zeta~B^2)~\overline{\rho^2}\simeq
\nu~,
\end{equation}
where $\nu=\frat{b s-4}{2}$. Apparently, the size distribution of pseudo-particles can be presented
by the well-known form as
\begin{equation}
\label{25}
\mu (\rho)= C_{N_c} \widetilde \beta^{2N_c}\Lambda^{b s} \rho^{b s-5}
 ~e^{-\nu~\frac{\rho^2}{\overline{\rho^2}}}~.
\end{equation}

Eqs. (\ref{22}) and (\ref{25}) allow us to get the estimate of generating functional (\ref{21}) in the following form
\begin{equation}
\label{26}
Y\ge \int D[B]~e^{-S(B)}~e^{-E}~,
\end{equation}
$$E=\int d^4x~
n~\left\{\ln\frat{n}{\Lambda^4}-1-\frat{\nu}{2}+\frat{\zeta~\overline{\rho^2}~B^2}{2}-
\ln \left[\frat{\Gamma(\nu)}{2}~C_{N_c}~\widetilde \beta^{2N_c}\right]-\nu~\ln
\frat{\overline{\rho^2}}{\nu} \right\}~.$$
Now taking into account Eq. (\ref{24}) and fixing a field $B$, parameters $s$ and $\lambda$ the maximum of
functional (\ref{26}) over the IL parameters can be calculated by solving the corresponding transcendental
equation ($\frac{d E}{d\bar\rho}=0$) numerically. Here it is a worthwhile place to notice the presence of
new factor in the denominator of $\frat{\Gamma(\nu)}{2}$ what is caused by the Gaussian form of the
corresponding integral over $\rho$ squared and, hence, the integration element requires the introduction
of $2\rho~d\rho$. In Ref. \cite{2} this factor was missed. However, this fact has not generated
a serious consequence because any application of these results is actually related to the choice of
suitable quantity of the parameter $\Lambda$ entering the observables (the pion decay constant, for example).
It means we should make the proper choice of basic scale. Besides, we should also keep in mind the approximate
character of IL model. Further we give the results for both versions to demonstrate the dependence of final
results on the renormalized constant $C_{N_c}$.

\begin{figure*}[!tbh]
\begin{center}
\includegraphics[width=0.5\textwidth]{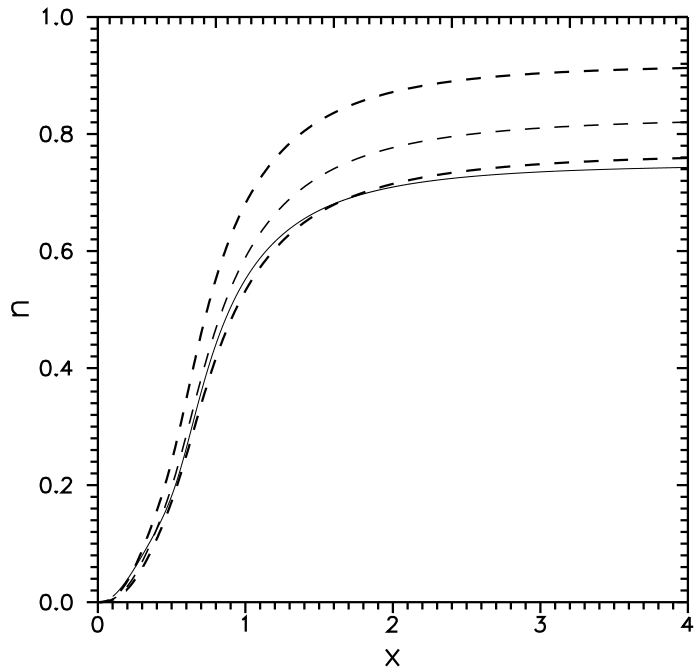}
\end{center}
  \vspace{-7mm}
 \caption{The IL density as the function of ${\mbox{x}}=y^2/\bar\rho^2$. Three dashed curves correspond
to the different profile functions. The lowest dashed line corresponds to the standard form (\ref{4}).
The top dashed line corresponds to the profile function with the screening factor (\ref{29})
and one parameter $\lambda$ ($s=1$) included and the middle line presents the same function but
with two parameters included. The solid line presents the selfconsistent solution of variational problem.}
  \label{fig2}
\end{figure*}

Searching the optimal configuration $f$ we take the effective action in the form of nonlinear functional as
\begin{equation}
\label{27}
S_{eff}=\int d^4x \left( \frat{G_{\mu\nu}^a(B)~
G_{\mu\nu}^a(B)}{4}+E[B] \right)~,
\end{equation}
in which the IL state is described by solutions $\bar\rho[B,s,\lambda]$, $n[B,s,\lambda]$. In practice the
following differential equation should be resolved
\begin{equation}
\label{28}
\frat{d^2 f}{d^2 y^2} =-\frat{1}{y^2}\frat{d f}{d y^2}+\frat{f(1-f)(1-2f)}{y^4}+\frat{1}{6\beta_0}~
\frat{d E}{d f}~,
\end{equation}
at fixed initial magnitude of $f({\mbox{x}}_0)$ putting up the derivative in the initial point $f'({\mbox{x}}_0)$
in such a way to have the solution going to zero when ${\mbox{x}}$ is going to infinity. Parameter $\beta_0$ is
introduced to fix a priori unknown value of coupling constant in the pseudo-particle definition (\ref{3}).
If the profile function has been fixed the configuration should be found in the form in which the starting values
of parameters $s$, $\lambda$ and $\beta_0$ coincide (within the given precision) with the  parameters
obtained from the solution $f$.
Nowadays this approach looks the most optimal one among other existing possibilities not only because of the
computational arguments but in view of the poor current level of understanding the interrelation between
perturbative and non-perturbative contributions
while calculating the effective Lagrangian. In fact, it was mentioned in Ref. \cite{2} that in more
general (realistic) formulation of this problem Eq. (\ref{28}) should include the term responsible for
the change of 'quantum' constant $C_{N_c}$ with the function $f$ changing. In principle, it could imply
that the problem of pseudo-particle ensemble stabilization is connected at the fundamental dynamics level
with the anticipated smallness of the $\frat{d C_{N_c}}{d f}$ contribution and, apparently, should be
addressed not so much to the description of the interacting pseudo-particles and their interactions
with the perturbative fields but rather to investigation of the time hierarchy corresponding to the
breakdown of quasi-stationary behaviour of the vacuum fluctuations which will certainly lead to the
changes of suitable effective Lagrangian (\ref{10}).

In order to receive the preliminary parameter estimates we consider the simplified model with the
profile function containing only one additional parameter for describing the screening effect as regards
\begin{equation}
\label{29}
f(y)=\frat{e^{-\alpha {\mbox{x}}}}{1+{\mbox{x}}}~,~~{\mbox{x}}=\frat{y^2}{\rho^2}.
\end{equation}
The energy $E$ as the function of the screening parameter $\alpha$ is depicted in Fig. 1.
The lowest dashed curve shows the behaviour when the changes related to weakening of repulsive interaction
are taken into account by switching on the parameter $\lambda$ only (at $s=1$). The top dashed curve was
obtained with both parameters switched on. The optimal value of the screening parameter $\alpha$
is determined by the minimum point of function $E(\alpha)$. Besides, this figure demonstrates the stability
of variational procedure of extracting the IL parameters. For the first calculation the values of
characteristic parameters for corresponding solution were taken as $\alpha=0.06$, $\lambda=0.775$,
$s=1.0067$ with the following set of the IL parameters $\bar\rho\Lambda=0.3305$,
$n/\Lambda^4=0.919$, $\beta=17.186$. These values give for the ratio of average pseudo-particle
size and average distance between pseudo-particles the quite suitable quantity $\bar\rho/R=0.324$.
For another calculation we have treated the parameter set characterizing the solution as $\alpha=0.02$,
$\lambda=0.888$, $s=1.0015$ and for the IL parameters the following values $\bar\rho\Lambda=0.315$,
$n/\Lambda^4=0.829$, $\beta=17.67$, $\bar\rho/R=0.3$. In order to get more orientation we would
like to mention that for the ensemble of standard pseudo-particles ($\alpha=0$, $\lambda=1$, $s=1$)
the corresponding values are $\bar\rho\Lambda=0.301$, $n/\Lambda^4=0.769$, $\beta=18.103$, $\bar\rho/R=0.282$.

\begin{figure*}[!tbh]
\begin{center}
\includegraphics[width=0.5\textwidth]{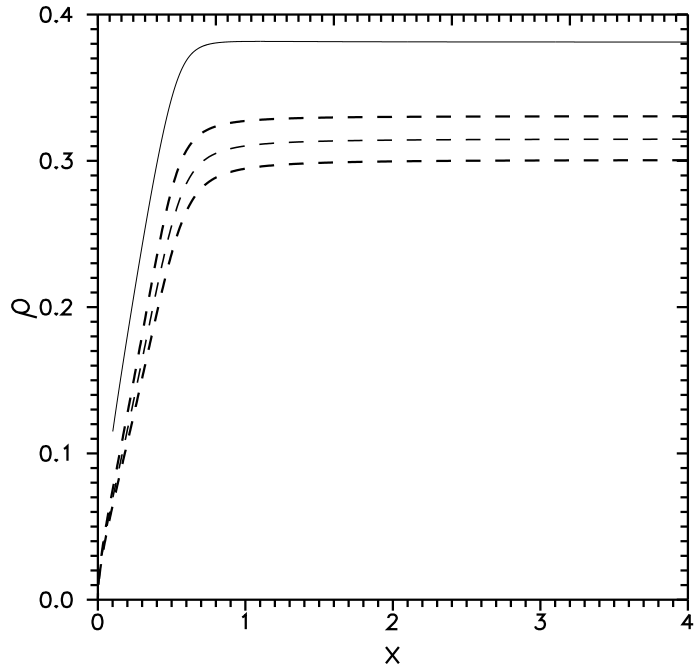}
\end{center}
  \vspace{-7mm}
 \caption{The average size of IL pseudo-particles as the function of ${\mbox{ x}}=y^2/\bar\rho^2$.
Three dashed curves correspond to different profile functions. The lowest curve corresponds to
the standard form (\ref{4}). The top dashed curve corresponds to the profile function with the
screening factor (\ref{29}) which includes one parameter $\lambda$ ($s=1$) and the middle line
shows the same function with two parameters included. The solid curve corresponds to the selfconsistent
solution of the variational problem.}
  \label{fig3}
\end{figure*}

Now we examine the impact of correction introduced in Eq. (\ref{26}) when we changed the term
$\frat{\Gamma(\nu)}{2}$ which has been obtained in Ref. \cite{2}. For the first calculation with
the set of solution parameters as $\alpha=0.24$, $\lambda=0.546$, $s=1.029$ we have for the IL
parameters $\bar\rho\Lambda=0.331$, $n/\Lambda^4=1.844$, $\beta=17.173$ which lead to the ratio
discussed equal to $\bar\rho/R=0.386$. For another calculation we have the following results
$\alpha=0.05$, $\lambda=0.799$, $s=1.0053$ and $\bar\rho\Lambda=0.291$, $n/\Lambda^4=1.356$,
$\beta=18.483$, $\bar\rho/R=0.314$. And for the ensemble of standard pseudo-particles
($\alpha=0$, $\lambda=1$, $s=1$) these parameters are
$\bar\rho\Lambda=0.265$, $n/\Lambda^4=1.186$, $\beta=19.305$, $\bar\rho/R=0.277$.

The Fig. 2 and Fig. 3 show the behaviours of IL density and average pseudo-particle size as
the functions of distance $x$. The dashed lines on both plots correspond to the similar ensembles.
The lowest curves demonstrate the behaviours for the ensembles of standard pseudo-particles (\ref{4}).
The top curves present the ensemble of pseudo-particles with the profile function
(\ref{29}) at $\alpha=0.06$ and $s=1$. And the middle dashed lines correspond to the profile functions
with $\alpha=0.02$ and $s\sim 1.03$. Obviously, it may be concluded that including even small change
of the second parameter value ($s\sim 1.03$) leads to the noticeable change of ensemble characteristics
(for example, the IL density) because the highest contribution to the action when the coupling constant
becomes the function of $\rho$ is essentially modified.

Let us make now several comments as to the 'complete' formulation of the problem of analyzing
the equation (\ref{28}). It was numerically resolved by the Runge-Kutta method. This approach
combined with numerical calculation of the derivative $\frat{d E}{d f}$ at every point of
consequent integration interval allows us to avoid the problems which appear when searching the
minimum of complicated functional in multidimensional space.

The initial data were fixed at the point ${\mbox{x}}_0=\frat{y_{0}^2}{\bar\rho^2}=0.1$.
Since the IL density value at the coordinate origin is inessential the initial form of
pseudo-particle profile function is taken without any deformations as
$f({\mbox{x}}_{0})=\frat{1}{1+{\mbox{x}}_{0}}$.
Then at fixed values of the parameters $\lambda$, $s$ and $\beta_0$ the coefficient $c$
is calculated. It allows to set the slope of trajectory
$f'({\mbox{x}}_{0})=-c f (1-f)/{\mbox{x}}_{0}$ at initial point in such a form in order to
have the solution going to zero at large distances. Afterwards we find out the values of
parameters $\lambda$ and $s$ requiring the input data to coincide with the output ones
within the fixed precision. The parameter values which obey the imposed constraints
are the following (input values) $\lambda=0.69099$, $s=1.049$, $\beta_0=16.26$ at
$c=1.361$ and $\lambda=0.691$,
$s=1.049$, $\beta_0=16.263$ (at the output of variational procedure). The solid line in Fig. 4
shows the obtained profile $f$ as the function of ${\mbox{x}}=\frat{y^2}{\bar\rho^2}$.
The differences of profiles are smoothed over if they are presented as the functions of
$y$ because the large magnitude of the screening coefficient, for example $\alpha=0.06$,
is compensated by enlargening the pseudo-particle size. The dashed lines on this plot show the
profile functions for the standard form (\ref{4}) (top dashed line), with the screening
factor (\ref{29}) including one parameter only $\alpha$ ($s=1$) (lowest dashed curve) and
two parameters included (middle dashed line).

\begin{figure*}[!tbh]
\begin{center}
\includegraphics[width=0.5\textwidth]{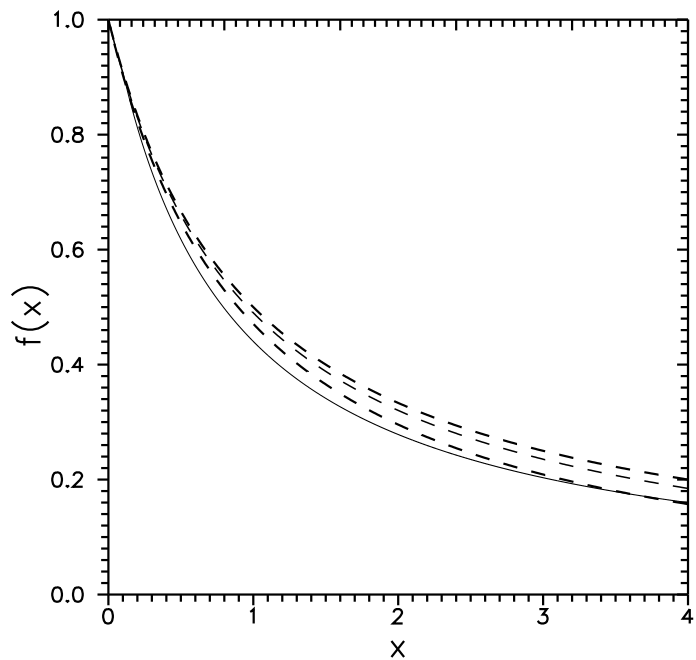}
\end{center}
  \vspace{-7mm}
 \caption{The various profile functions. The top dashed curve corresponds to the standard form (\ref{4}),
the lowest dashed curve shows the function with the screening factor (\ref{29}) including one parameter
$\lambda$ ($s=1$) and the middle line presents the same function with two parameters included.
The solid line corresponds to the selconsistent solution of variational problem.}
  \label{fig4}
\end{figure*}

Another calculation (with modified $\Gamma$-function contribution) was based on the slightly different set of
relevant parameters which are for the input values
$\lambda=0.607$, $s=1.0515$, $\beta_0=17.04$ at $c=1.545$ and
$\lambda=0.6066$, $s=1.0515$, $\beta_0=17.042$ for the output one
at the finish of variational procedure. The behaviours of IL
density and average pseudo-particle size for selfconsistent
solution are plotted in Fig. 2 and Fig. 3 (solid lines,
respectively){\footnote{It is interesting to notice that
considering IL (ensemble of pseudo-particles in the singular
gauge) in the field of regular pseudo-particle we obtain the IL
density value in the center of regular pseudo-particle which is
larger than its value at large distances what looks like the
anti-screening effect.}}. In the Table 1 we present the IL
parameters at the large distances from pseudo-particle (the first
line) together with the data for the ensemble of pseudo-particles
with the standard profile function (the second line). The third
and fourth lines of this Table 1 are devoted to the calculations
with the second set of parameters (with factor $2$ absent in Eq.
(\ref{26})). The fourth line, in particular, presents the
calculations for pseudo-particles with standard form of profile
function.
\begin{center}
{\underline{ Table 1}}. Parameters of IL.
\\\vspace{0.3cm}
\begin{tabular}{|c|c|c|c|c|c|}
\hline
$\bar\rho\Lambda$&$n/\Lambda^4$&$\beta$ &$\bar\rho/R$&$n\bar\rho^4$ \\\hline
$0.381$          &$0.743$       &$16.263$&$0.354$     &$\dmn{1.582}{-2}$ \\
$0.331$          &$0.769$       &$18.103$&$0.282$     &$\dmn{6.277}{3}$ \\\hline\hline
$0.354$          &$1.245$       &$17.042$&$0.379$     &$\dmn{1.955}{-2}$ \\
$0.265$          &$1.186$       &$19.305$&$0.277$     &$\dmn{5.849}{-3}$ \\
\hline
\end{tabular}
\end{center}
It is quite obvious that the utilization of optimal
pseudo-particle profile function leads to the larger
pseudo-particle size but the packing fraction parameter holds,
nevertheless, a small quantity which is quite suitable for the
perturbative expansion. Besides, the results obtained allow us to
conclude that with tuning $\Lambda$ a fully satisfactory agreement
our calculations of pseudo-particle size, the ensemble diluteness
and gluon condensate value with their phenomenological magnitudes
extracted from the other models are easily reachable. The
calculations of several dimensional quantities in our approach are
also very indicative. The values of the screening mass (\ref{11}),
average pseudo-particle size and IL density obtained for two
values of $\Lambda$ ($200$ MeV and $280$ MeV) are shown in Table
2. The sequence of line meanings is identical to that in Table 1
as well as the meanings of last four lines which present the
results of calculations with the second set of parameters (with
factor $2$ absent in Eq. (\ref{26})).
\begin{center}
{\underline{ Table 2}}. Screening mass and IL parameters
\\\vspace{0.3cm}
\begin{tabular}{|c|c|c|c|c|c|}
\hline
$\Lambda$~MeV&$m$~MeV&$\bar\rho$~GeV$^{-1}$&$n$ fm$^{-4}$ \\\hline
$200.$       &$381$ &$1.906$               &$0.7496$ \\
             &$304$ &$1.503$               &$0.7688$ \\
\hline
$280.$       &$533$ &$1.361$               &$2.88$ \\
             &$426$ &$1.074$               &$2.95$ \\
\hline\hline
$200.$       &$456$ &$1.77$                &$1.245$ \\
             &$333$ &$1.325$               &$1.186$ \\
\hline
$280.$       &$638$ &$1.264$               &$4.78$ \\
             &$466$ &$0.946$               &$4.56$ \\
\hline
\end{tabular}
\end{center}
Another interesting feature of this calculation is the weakening of pseudo-particle interaction.
This effect is driven by the coefficient $\xi^2$ ($\sim \lambda^2$). Our estimates for the first
set of parameters give $\lambda=0.691$ and, hence, $\lambda^2\sim 0.48$ and for the second set
we have ($\lambda=0.607$) and $\lambda^2\sim 0.37$. Let us mention here that the reasonable
description of instanton ensemble can be reached in the framework of two-component models
\cite{6} as well.

Our calculations enable us to conclude that dealing with IL model
(formulated in one-loop approach) one is able to reach quite reasonable description of gluon
condensate even being constrained by the values of average pseudo-particle size and other
routine phenomenological parameters. Moreover, the ensemble of pseudo-particles with standard
profile functions turns out to be very practical because introducing the other configurations
to make the similar estimates is simply unoperable. With such an approximation of the vacuum
configurations the coefficient of interaction weakening develops the magnitude about
$\lambda^2\sim 0.3$ --- $0.5$. Including this effect leads to the enlargening of pseudo-particle size.
It allows us to conclude that nowadays the instantons in the singular gauge is the only
serious instrument for effective practising.

The authors are sincerely grateful to  A.E. Dorokhov and S.B. Gerasimov for interesting
discussions and practical remarks. The financial support of the Grants INTAS-04-84-398
and NATO PDD(CP)-NUKR980668 is also acknowledged.


\end{document}